\documentclass[a4paper, 12pt]{article}
\usepackage[top=25.4mm,bottom=25.4mm,left=25.4mm,right=25.4mm]{geometry} % Document margins
\usepackage{setspace}
\usepackage[english]{babel}
\usepackage[utf8]{inputenc}
\usepackage[para,online,flushleft]{threeparttable}
\usepackage{lscape}
\usepackage{fixltx2e}
\usepackage{amsmath}
\usepackage{placeins}
\usepackage{amsfonts}
\usepackage{amssymb}
\newcommand{\sym}[1]{\rlap{#1}}
\usepackage{makeidx}
\usepackage{epsfig}
\usepackage{comment}
\usepackage{amsthm}
\usepackage{verbatim}

\usepackage{graphicx}
\usepackage{morefloats}
\usepackage{rotfloat}
\usepackage{adjustbox}
\usepackage{multirow}
\usepackage{lipsum}
\usepackage{eqnarray}

\usepackage{moreverb}
\usepackage{longtable}
\usepackage{url}
\usepackage{breakurl}

\usepackage[breaklinks]{hyperref}

\usepackage{keyval}
\usepackage{ifthen}
\usepackage{etex}

\usepackage{csquotes}
\usepackage[backend=biber,style=authoryear,maxcitenames=2,maxbibnames=50,firstinits=true,useprefix=true,dashed=false]{biblatex}
\DeclareNameAlias{sortname}{last-first}
\renewbibmacro{in:}{%
 \ifentrytype{article}{}{\printtext{\bibstring{in}\intitlepunct}}}

\usepackage{xpatch}
\xpatchbibmacro{journal+issuetitle}{%
  \setunit*{\addspace}%
  \iffieldundef{series}}
  {%
  \setunit*{\addcomma\space}%
  \iffieldundef{series}}{}{}

\bibliography{SC_input.bib}

\renewbibmacro*{volume+number+eid}{%
  \printfield{volume}
  \printfield[parens]{number}%
  \setunit{\addcomma\space}%
  \printfield{eid}}

\usepackage{graphicx}
\usepackage{subcaption}
\usepackage{amsmath, amsthm, amssymb}
\usepackage{stackengine}
\usepackage{verbatim}
\usepackage{txfonts}
\usepackage{etoolbox}
\usepackage{ifxetex}
\usepackage{pgfopts}
\usepackage{calc}
\usepackage[symbol]{footmisc}

\usepackage{eurosym}

\usepackage{xcolor}
\usepackage{pgffor}
\usepackage[version=3]{mhchem}
\usepackage{float}
\usepackage{import}
\usepackage{booktabs}
\usepackage{tabularx}
\usepackage{tikz}

\usetikzlibrary{patterns}

\usepackage{pgfplots}
\usepgfplotslibrary{polar}
\usetikzlibrary{calc}
\usetikzlibrary{snakes}
\usepackage{enumitem}
\usepackage{framed}
\usepackage[toc,page]{appendix}

\usepackage{placeins}
\usepackage{titling}
\newcommand{\logo}[1]{%
  \postauthor{%
  \end{tabular}\par\end{center}
  \begin{center}
  \vskip8.0em
  \includegraphics[scale=0.22]{#1}
  \end{center}
  \vskip4.0em}%
}
	\makeatletter
	\g@addto@macro{\figure}{\centering}
	\makeatother
	\makeatletter
	\g@addto@macro{\table}{\centering}
	\makeatother

\usepackage{lmodern}
\usepackage[T1]{fontenc}
\usepackage{avant}
\usepackage[labelfont=bf]{caption} %make table labels bold

\renewcommand{\thefootnote}{\fnsymbol{footnote}}
\begin{document}

\thispagestyle{empty}

\begin{center}

\vspace*{2cm}

{\LARGE Are temporary value-added tax reductions passed on to consumers? Evidence from Germany's stimulus}

\vspace{0.75cm}

{\large Felix Montag, Alina Sagimuldina and Monika Schnitzer\footnote[1]{Department of Economics, Ludwig-Maximilians-University Munich. E-mail addresses for correspondence: felix.montag@econ.lmu.de; alina.sagimuldina@econ.lmu.de; schnitzer@econ.lmu.de. Financial support by the Deutsche Forschungsgemeinschaft through CRC TRR 190 is gratefully acknowledged. Montag thanks the International Doctoral Program ``Evidence-Based Economics'' of the Elite Network of Bavaria for financial support. Sagimuldina thanks the Research Training Group ``GRK 1928 - Microeconomic Determinants of Labor Productivity'' of the Deutsche Forschungsgemeinschaft for financial support.}}

\vspace{0.75cm}

{\large \today}
\end{center}

\thispagestyle{empty}

\vspace{1.0cm}

\begin{singlespacing}
\begin{abstract}

This paper provides the first estimates of the pass-through rate of the ongoing temporary value-added tax (VAT) reduction, which is part of the German fiscal response to COVID-19. Using a unique dataset containing the universe of price changes at fuel stations in Germany and France in June and July 2020, we employ a difference-in-differences strategy and find that pass-through is fast and substantial but remains incomplete for all fuel types. Furthermore, we find a high degree of heterogeneity between the pass-through estimates for different fuel types. Our results are consistent with the interpretation that pass-through rates are higher for customer groups who are more likely to exert competitive pressure by shopping for lower prices. Our results have important implications for the effectiveness of the stimulus measure and the cost-effective design of unconventional fiscal policy.

\end{abstract}

\vspace{0.75cm}

\noindent

{\small \medskip }


\vspace{0.5cm}


\vspace{3.25cm}

\end{singlespacing}
\pagebreak

%%%%%%%%%%%%%%%%%%%%%%%%%%%%%%%%%%%%%%%%%%%%%%%%%%%%%%%%%%%%%%%%%%%%%%%%%%%%%%%%%%%%%%%%%%%%%%%
%\doublespacing
\renewcommand*{\thefootnote}{\arabic{footnote}}
\setcounter{footnote}{0}
\setcounter{page}{1}
\pagenumbering{arabic}
%\newpage
%\tableofcontents
%\newpage
%this is scc_intro

\section{Introduction}

The drastic economic downturn accompanying the COVID-19 pandemic was met by an unprecedented fiscal response of the German government. On 3 June 2020, a stimulus package worth 130 billion Euro was announced. To the general surprise of the public, it included a reduction of the standard value-added tax (VAT) rate from 19 to 16 percent and of the reduced rate from 7 to 5 percent for the second half of 2020, at an estimated cost of 20 billion Euro or 0.6 percent of GDP. The aim of this fiscal policy is to temporarily reduce prices and stimulate consumption through inflation expectations. For this to work, however, it is crucial that firms pass on the VAT reduction to consumers.

Our paper provides the first estimates of the pass-through rate of the temporary VAT reduction for a major sector of the economy. We estimate the pass-through rate for diesel and gasoline using a unique dataset containing the universe of price changes at fuel stations in Germany and France in June and July 2020 and employing a difference-in-differences strategy. We find that pass-through is incomplete for all types of fuels and very heterogeneous across fuel types. This variation of pass-through rates is consistent with the interpretation that customers of different fuel types are differentially likely to shop for lower prices, resulting in differing competitive pressure.  Whilst fuel stations pass on most of the VAT rate reduction to diesel customers who on average drive more than twice as many kilometers per year than drivers of gasoline driven cars, pass-through rates for gasoline are much lower.

Studying the effect of the temporary VAT rate reduction in the context of the fuel market is particularly interesting for two reasons. First, it is a market where granular data is available in real-time, which allows us to evaluate the effect while it is happening. Second, it is a market where price adjustments are costless and in fact happen frequently. Thus, despite the temporary nature of the VAT rate change, price adjustment costs cannot be held accountable for imperfect pass-through.

This paper makes two contributions to the literature. First, our results have implications for the effective design of unconventional fiscal policy. \textcite{feldstein2002} proposed that stimulating inflation in an environment where monetary policy is ineffective could be done by targeting household expectations directly. This type of policy was later coined unconventional fiscal policy by \textcite{dacunto2018}. A temporary reduction in the VAT rate is only likely to affect household expectations, however, if the reduction is passed on to consumers and they therefore expect prices to be lower temporarily. By analyzing the pass-through rate of the temporary VAT reduction, we shed light on a necessary condition for this type of unconventional fiscal policy to be effective. Furthermore, we show that by targeting competitive markets where consumers are likely to search for lower prices, policymakers can make unconventional fiscal policy more cost-effective. \textcite{dacunto2020} find that a previous instance of unconventional fiscal policy in Germany, an increase in the VAT rate from 16 to 19 percent announced in 2005 and implemented in 2007, increased inflation expectations and durable expenditure. Based on this evidence, \textcite{dacunto2020b} predict that the 2020 temporary VAT reduction will successfully increase inflation expectations and expenditure. \textcite{benzarti2020} show that pass-through is often asymmetric and that prices respond twice as much to VAT increases as to decreases. Furthermore, whilst the 2007 VAT increase was permanent, the current VAT change only lasts for six months. Conclusions from the 2007 permanent VAT increase therefore are not necessarily informative about the 2020 temporary VAT reduction.

Second, we are the first to provide estimates of the average pass-through rate of an unanticipated VAT rate change in a major sector of the economy using high-frequency, establishment-level price data. Our unique dataset allows us to observe all price changes for around $23,000$ fuel stations across Germany and France before and after the temporary VAT rate reduction. Fuel stations in Germany are treated after the 1 July, whereas fuel stations in France are unaffected by the German policy change. Thus, we can employ a difference-in-differences strategy, using French fuel stations as control group. Previous empirical studies on tax pass-through use aggregate price indices, finding mixed results. They include evidence for under-shifting (e.g. \cite{benzarti2019}), full pass-through (e.g. \cite{benedek2019}) or over-shifting (e.g. \cite{besley1999}). Notable exceptions using firm-level price data are \textcite{kosonen2015}, studying the effect of a VAT reduction for hairdressers in Finland, and \textcite{buettner2020}, who estimate the effect of a large number of VAT increases and decreases in the European Union (EU) using product-level monthly sales data for home appliances in a panel model across a large number of European countries over time. Like \textcite{kosonen2015}, we focus on the effect of one particular policy change on one particular market. In contrast to \textcite{kosonen2015}, using a control market in a different country allows us to avoid potential general equilibrium effects affecting the control group, which might lead to an over- or under-estimation of the pass-through rate, as noted by \textcite{benedek2019}. Finally, a feature of high-frequency price data in a market with many price adjustments is that it allows us to trace out the evolution of pass-through rates over time.

To estimate the average pass-through rate of the VAT reduction, we use a difference-in-differences strategy, where we compare daily prices of the three main fuel types sold at fuel stations in Germany and France before and after the policy change.\footnote{This dataset has previously been used by \textcite{montag2020} to analyze the effect of price transparency.} Supply shocks, in particular fluctuations in the price of crude oil, should affect Germany and France similarly and are thus eliminated by time fixed effects. We also account for regional differences in demand over time by controlling for changes in mobility, using regional data from the Google COVID-19 Community Mobility Report. 

We find that the pass-through rate for diesel is 83 percent, whereas it is 61 percent for \textit{E10} and 40 percent for \textit{E5}.\footnote{\textit{E5} and \textit{E10} are the two main types of gasoline sold in Germany, which differ in how much bioethanol they contain.} This translates into price decreases of 2 percent for diesel, $1.5$ percent for \textit{E5} and 1 percent for \textit{E10}. At the same time, retail margins for diesel only increased by 0.7 percent, whilst retail margins for \textit{E5} and \textit{E10} increased by between 10 and 12 percent.\footnote{Our measure of retail margins only subtracts taxes, duties and the price of crude oil at the port of Rotterdam from fuel prices. It includes fuel station and refinery margins, as well as different cost types. The percentage change in retail margins due to the VAT change is therefore an underestimate of the actual percentage change in retail margins.}

Our results show that pass-through of the VAT rate reduction is fast and substantial but remains incomplete. Whilst prices decrease for consumers, margins also substantially increase for sellers. Furthermore, there is a substantial difference in pass-through rates between fuel types. Since stations sell all three types of fuel, unobserved station characteristics cannot explain these differences. Instead, differences in competitive pressure due to different propensities of customer groups to shop for lower prices are consistent with the observed effects. Whereas fuel stations pass on more than half of the VAT rate decrease to frequent drivers and professional drivers, usually driving diesel cars, as well as to price sensitive gasoline customers who buy the cheaper \textit{E10}, they pass on less than half of the tax rate reduction to customers of \textit{E5} suggesting that they are less responsive to price differentials.

The remainder of the paper is structured as follows: Section \ref{sec: industry} describes the industry, Section \ref{sec: data} gives an overview of the data and presents descriptive evidence, Section \ref{sec: empirical} discusses the empirical strategy, Section \ref{sec: results} presents the estimation results and Section \ref{sec: conclusion} concludes.

%\input{scc_policy}
%this is scc_policy

\section{The Retail Fuel Market}\label{sec: industry}

%% The importance
In 2019, total revenues from retail fuel sales were worth 92 billion Euro or approximately 3 percent of German GDP. In addition to its standalone value to the economy, this market has large externalities on the rest of the economy. Fuel prices are a key determinant of travel costs, commuting costs and, more broadly, the cost of personal transportation.

%% The products
The first important distinction to make within fuels for passenger vehicles is between diesel and gasoline.\footnote{Since fuel stations do not report prices for truck diesel to the Market Transparency Unit, we only focus on fuel prices for passenger vehicles.} In Germany, diesel has a volume share of 44 percent of fuel for passenger vehicles with combustion engines and gasoline accounts for the remaining 56 percent.\footnote{This is based on 2018 figures from \textit{Verkehr in Zahlen 2019/2020}, published by the Federal Ministry of Transportation. To the best of our knowledge, these are the most recent administrative figures concerning the passenger vehicle market only.} Substituting between these two types of fuel is very costly, both on the demand and supply side.\footnote{On the demand side, this would usually require buying a new vehicle. On the supply side, readjusting the ratio of diesel and gasoline made from a barrel of crude oil is possible, but only to a limited extent and at a high cost.} Within gasoline, there is differentiation according to the octane rating and the share of ethanol. Standard gasoline (commonly referred to as \textit{Super}) has an octane rating of 95. It has a volume share of $95.4$ percent of the gasoline market.\footnote{This is based on 2019 figures from the monthly oil statistics, published by the Federal Office for Economic Affairs and Export Control.} Some high-perfomance vehicles will require gasoline with an octane rating of 98 (commonly referred to as \textit{Super Plus}), which has a $4.6$ percent volume share within gasoline. At the same time, there is no added benefit of fueling \textit{Super Plus} if the vehicle can process \textit{Super}. Since the price of \textit{Super Plus} is always considerably higher, there is no demand-side substitution from \textit{Super} to \textit{Super Plus} either. As fuel stations do not report prices of \textit{Super Plus} to the Market Transparency Unit in Germany, it is not part of our analysis.
% Diesel & gasoline shares: Verkehr in Zahlen 2019/2020, p. 309.
% Super & Super Plus shares: Amtliche Mineralölstatistik, Dezember 2019, Tabelle 6c), Fußnote d).

Within \textit{Super}, there is a final distinction according to the share of ethanol. Standard gasoline has a 5 percent share of ethanol and is thus commonly referred to as \textit{E5}. In 2011, a new type of gasoline was introduced with a 10 percent ethanol share, referred to as \textit{E10}. The aim of increasing the share of ethanol is to reduce greenhouse gas emissions and decrease the amount of fossil fuel used in transportation. Although \textit{E5} and \textit{E10} are not taxed differently, \textit{E10} is usually around 4 Eurocent cheaper than \textit{E5}. This is partly driven by the relative prices of crude oil and ethanol on the world market and partly by a minimum quota of biofuels that need to be sold by fuel stations.

After the introduction of \textit{E10} in 2011, there was controversy about whether biofuels damage the engine. Although biofuels can pose a significant threat to the engine of a vehicle that is not certified to be compatible with \textit{E10}, around 90 percent of gasoline-run vehicles, including all vehicles produced after 2012, are compatible with \textit{E10}.\footnote{A full list of compatible vehicles can be found here \url{https://www.dat.de/e10/}.} According to the German Automobile Association, \textit{E10} is around $1.5$ percent less efficient than \textit{E5}.\footnote{See \url{https://www.adac.de/verkehr/tanken-kraftstoff-antrieb/benzin-und-diesel/e10-tanken/}.} All fuel stations in Germany are required to sell both types of fuel. Nevertheless, in 2019 \textit{E5} still had a volume share of $85.6$ percent within \textit{Super} and \textit{E10} only of $14.4$ percent. Overall, many motorists who could buy less expensive \textit{E10} choose not to do so and buy \textit{E5} instead. Reasons for this could include preferences or a lack of information, which point towards a lower price sensitivity of \textit{E5} customers compared to \textit{E10} customers.
% E5 & E 10 shares: Amtliche Mineralölstatistik, Dezember 2019, Tabelle 6c), Fußnote d).

Two further observations can be made about the difference between drivers of gasoline and diesel passenger vehicles. Whereas only 32 percent of registered passenger vehicles in Germany have a diesel engine, compared to 66 percent that run on gasoline, frequent drivers tend to use diesel cars.\footnote{This is based on April 2020 figures on registered passenger vehicles in Germany, published by the German Federal Motor Transport Authority.} On average, gasoline passenger vehicles drive $10,800$ kilometers, whereas diesel passenger vehicles drive $19,500$ kilometers per year.\footnote{This is based on 2018 figures from \textit{Verkehr in Zahlen 2019/2020}, published by the Federal Ministry of Transportation.}
% Diesel & gasoline average kilometers: Verkehr in Zahlen 2019/2020, p. 309.

%% Cost structure
The largest share of the fuel price consists of taxes. A lump-sum energy tax of 0.6545 Euro per liter is levied on gasoline (0.4704 Euro per liter for diesel).\footnote{An additional fuel storage fee of 0.0027 Euro per liter is levied on gasoline and 0.0030 Euro per liter on diesel.} In addition, there is a 19 percent value-added tax which is levied on the net price of diesel and gasoline, including the energy tax. This value-added tax is temporarily reduced to 16 percent between July and December 2020.
% Taxes and levies: https://www.avd.de/kraftstoff/staatlicher-anteil-an-den-krafstoffkosten/

Crude oil accounts for another important share of the fuel price and is the most important source of price fluctuations. A barrel (42 gallons) of crude oil can be refined into around 19 gallons of gasoline, 12 gallons of diesel, as well as 13 gallons of other products, such as jet fuel, petroleum coke, bitumen or lubricants.\footnote{These are approximate shares which can vary by context and type of crude oil. The total volume of products refineries produce (output) is greater than the volume of crude oil that refineries process (input) because most of the products they make have a lower density than the crude oil they process. See \url{https://www.eia.gov/energyexplained/oil-and-petroleum-products/refining-crude-oil-inputs-and-outputs.php}.} Gasoline and diesel are the most valuable components of refined crude oil.
% Refining details: https://www.eia.gov/energyexplained/oil-and-petroleum-products/refining-crude-oil-inputs-and-outputs.php

%this is scc_data 		

\section{Data and Descriptive Evidence}\label{sec: data}

Our dataset contains all price changes for close to all fuel stations in Germany and France, as well as several characteristics of these stations.\footnote{In Germany, only gasoline stations at highway service areas are exempt from reporting their price changes. In France, fuel stations selling less than $500 m^{3}$ per year are exempt from reporting price changes.} In Germany, stations report price changes in real-time to the Market Transparency Unit at the German Federal Cartel Office. Tankerkönig, a price comparison website, provides access to this data, as well as to station characteristics, to researchers.\footnote{See \url{https://creativecommons.tankerkoenig.de/}.} Similarly, price changes in France have to be reported by stations to a government agency, which makes this data available to researchers.\footnote{See \url{https://www.prix-carburants.gouv.fr/rubrique/opendata/}.} Furthermore, we add data on the daily price of crude oil, the principal input product for diesel and gasoline, at the port of Rotterdam. Finally, we use data on daily regional mobility patterns from the COVID-19 Community Mobility Report provided by Google.

Our analysis starts on 15 June 2020 and currently goes until 31 July 2020. The 15 June is a natural starting point, as it marks the beginning of the European Commission's \textit{Re-open EU} plan and is the date on which France and Germany lifted many of their travel restrictions and re-opened their borders.\footnote{See \url{https://reopen.europa.eu/en/}.} As long as the temporary VAT rate reduction remains in place, we plan to extend our period of analysis and estimate how the pass-through rates evolve.

Using the data on price changes, we construct daily weighted average prices. Table \ref{tab: summary} shows that the price level is generally higher in France than in Germany. Gross prices in France increase by around 2 Eurocent between the pre- and post-VAT cut periods. In Germany, gross prices remain constant for \textit{E5} and \textit{E10} and decrease by 1 Eurocent for diesel. At the same time, the increase in the net price in Germany is between 2 and 3 Eurocent, depending on the fuel type, which is larger than in France and thus suggests that the VAT reduction was not completely passed on to consumers.

\FloatBarrier

\begin{table}\centering
\def\sym#1{\ifmmode^{#1}\else\(^{#1}\)\fi}
\begin{threeparttable}
\caption{Summary statistics \label{tab: summary}}
\begin{footnotesize}
\begin{tabular}{lcccc}
\toprule

\multicolumn{1}{l}{} &\multicolumn{1}{c}{Germany} &\multicolumn{1}{c}{Germany} &\multicolumn{1}{c}{France} &\multicolumn{1}{c}{France} \\ 
\multicolumn{1}{l}{} &\multicolumn{1}{c}{pre-VAT cut} &\multicolumn{1}{c}{post-VAT cut} &\multicolumn{1}{c}{pre-VAT cut} &\multicolumn{1}{c}{post-VAT cut} \\ \midrule
\multicolumn{5}{l}{A. Station characteristics}                       \\
\multicolumn{1}{l}{Number of stations}                   &\multicolumn{1}{c}{14,872}                                 &\multicolumn{1}{c}{14,939}                                &\multicolumn{1}{c}{8,760}                                      &\multicolumn{1}{c}{8,829}                                       \\
%\multicolumn{1}{l}{Share of integrated stations}         &\multicolumn{1}{c}{55\%}                                  &\multicolumn{1}{c}{55\%}                                   &\multicolumn{1}{c}{}                                           &\multicolumn{1}{c}{}                                            \\
 \midrule
\multicolumn{5}{l}{B. Prices and margins, \textit{E5}}                               \\
\multicolumn{1}{l}{Mean price}                           &\multicolumn{1}{c}{1.28}                                  &\multicolumn{1}{c}{1.28}                                   &\multicolumn{1}{c}{1.34}                                      &\multicolumn{1}{c}{1.36}                                       \\
\multicolumn{1}{l}{Mean price net of taxes and duties}   &\multicolumn{1}{c}{.41}                                   &\multicolumn{1}{c}{.44}                                    &\multicolumn{1}{c}{.43}                                       &\multicolumn{1}{c}{.44}                                        \\
\multicolumn{1}{l}{Mean retail margin}                   &\multicolumn{1}{c}{.14}                                   &\multicolumn{1}{c}{.17}                                    &\multicolumn{1}{c}{.16}                                       &\multicolumn{1}{c}{.16}                                        \\ \midrule
\multicolumn{5}{l}{C. Prices and margins, \textit{E10}}           \\
\multicolumn{1}{l}{Mean price}                           &\multicolumn{1}{c}{1.24}                                  &\multicolumn{1}{c}{1.24}                                   &\multicolumn{1}{c}{1.31}                                      &\multicolumn{1}{c}{1.33}                                       \\
\multicolumn{1}{l}{Mean price net of taxes and duties}   &\multicolumn{1}{c}{.39}                                   &\multicolumn{1}{c}{.41}                                    &\multicolumn{1}{c}{.42}                                       &\multicolumn{1}{c}{.44}                                        \\
\multicolumn{1}{l}{Mean retail margin}                   &\multicolumn{1}{c}{.12}                                   &\multicolumn{1}{c}{.13}                                    &\multicolumn{1}{c}{.15}                                       &\multicolumn{1}{c}{.16}                                        \\ \midrule
\multicolumn{5}{l}{D. Prices and margins, diesel}        \\
\multicolumn{1}{l}{Mean price}                           &\multicolumn{1}{c}{1.08}                                  &\multicolumn{1}{c}{1.07}                                   &\multicolumn{1}{c}{1.23}                                      &\multicolumn{1}{c}{1.25}                                       \\
\multicolumn{1}{l}{Mean price net of taxes and duties}   &\multicolumn{1}{c}{.43}                                   &\multicolumn{1}{c}{.45}                                    &\multicolumn{1}{c}{.42}                                       &\multicolumn{1}{c}{.43}                                        \\
\multicolumn{1}{l}{Mean retail margin}                   &\multicolumn{1}{c}{.16}                                   &\multicolumn{1}{c}{.17}                                    &\multicolumn{1}{c}{.15}                                       &\multicolumn{1}{c}{.15}                                        \\ 
\midrule
\multicolumn{5}{l}{E. Mobility data}                               \\
\multicolumn{1}{l}{Retail \& recreation}                           &\multicolumn{1}{c}{-9.3\%}                                  &\multicolumn{1}{c}{-2.8\%}                                   &\multicolumn{1}{c}{-9.5\%}                                      &\multicolumn{1}{c}{2.1\%}                                       \\
\multicolumn{1}{l}{Workplaces}   &\multicolumn{1}{c}{-14.7\%}                                   &\multicolumn{1}{c}{-19.8\%}                                    &\multicolumn{1}{c}{-16\%}                                       &\multicolumn{1}{c}{-22.8\%}                                        \\
\bottomrule
\end{tabular}
\end{footnotesize}
\begin{tablenotes}
\footnotesize Notes: ``pre-VAT cut'' and ``post-VAT cut'' refer to fuel stations in Germany and France before and after the reduction of the VAT rate, respectively. The pre-VAT phase goes from 15 June until 31 June 2020. The post-VAT phase starts on 1 July 2020.\\
\end{tablenotes}
\end{threeparttable}
\end{table}

\FloatBarrier

We also calculate retail margins by subtracting taxes, duties and the share of the price of crude oil that goes into the production of diesel and gasoline, respectively.\footnote{For a detailed description of the calculation of prices and margins, see Appendix \ref{app sec: data}.} Although these retail margins still contain different cost types, such as the cost of refining or transportation costs, the main source of input cost variation, the price of crude oil, is eliminated. Table \ref{tab: summary} shows that retail margins remained mostly constant for France before and after the VAT reduction. Although there is only a modest increase in retail margins for diesel in Germany after the VAT reduction, there is an increase in the retail margin of around 15 percent for \textit{E10} and 16 percent for \textit{E5}.\footnote{Percentage changes are different from what you would calculate from the retail margins in the table because of rounding of margins in the table.}

To capture regional changes in demand over time, we use the daily percentage change in visits to retail and recreation, as well as to the workplace, from the COVID-19 Community Mobility Report. With the former, we intend to capture local changes in the propensity of using a car for leisurly activities, including going on vacation. With the latter, we aim to capture local changes in the propensity to use a car for professional activities. Both of these variables are measured as the percentage change of activities compared to the median value for the corresponding day of the week during the five-week period 3 January to 6 February 2020. The data is disaggregated for 96 sub-regions in France and 16 regions in Germany. We use the geolocation of each fuel station to match the measures of local mobility to each station.

Table \ref{tab: summary} shows that mobility patterns in France and Germany are similar. Whereas visits to retail and recreational facilities were around 10 percent lower in the second half of June compared to the baseline beginning of the year, in July, the number of such visits returned close to their pre-pandemic levels. At the same time, in both countries visits to workplaces were around 15 percent lower in the second half of June compared to the baseline and 20 percent lower in July. This is likely because many people go on vacation in July. It also indicates that overall trends in both countries are very similar.

\section{VAT Pass-through Estimation}\label{sec: empirical}

We begin by estimating the average effect of the VAT reduction on fuel prices and retail margins in Germany. To do this, we compare the evolution of prices and retail margins at fuel stations in Germany to stations in France, before and after the decline in the value-added tax.

\subsection{Setup}

For the three main types of fuel, we focus on two main outcomes: daily fuel prices and retail margins. To causally estimate the effect of the temporary reduction in the VAT rate on fuel prices and retail margins, we use a difference-in-differences strategy, and compare stations in Germany and France, before and after the reduction in the VAT rate. Specifically, we estimate the following regression:

\begin{equation}\label{eq: did}
    Y_{it} = \beta_{0} + \beta_{1}VAT_{it} + \alpha X_{it} + \mu_{i} + \gamma_{t} + \epsilon_{it}
\end{equation}

where $Y_{it}$ is the logarithm of the price or retail margin of gasoline or diesel at a fuel station $i$ at date $t$, and $VAT_{it}$ is a dummy variable that equals one for stations affected by the VAT reduction at date $t$. These are fuel stations in Germany from 1 July 2020 onwards. $X_{it}$ is a vector of controls, which includes regional mobility data for retail and recreational purposes, and mobility to work. $\mu_{i}$ and $\gamma_{t}$ correspond to fuel station and date fixed effects, respectively.

\subsection{France as a Control Group}

To causally identify the effect of the VAT reduction on fuel prices and retail margins, two main assumptions must be satisfied. First, there should be no transitory shocks that would differentially affect fuel stations in Germany and France before and after the reduction in VAT, other than the policy change itself. Second, there should be no spillover effects from the VAT reduction in Germany onto the fuel market in France.

Station fixed effects control for any time-invariant differences between fuel stations in France and Germany, and date fixed effects capture the transitory shocks, such as fluctuations in the price of crude oil, that identically affect French and German stations. The two countries are similar in their geographic location, size, and wealth. Since in our analysis we also focus on a relatively narrow window around the reform, this should alleviate concerns on transitory shocks differentially affecting French and German fuel stations.

One might still suspect that certain transitory shocks could confound our empirical strategy. We now discuss the most obvious candidates. On the demand side, public and school holidays in France and Germany are highly correlated. Travel restrictions put in place due to COVID-19 were lifted simultaneously in the two countries. Starting from 15 June 2020, residents of the Schengen Area and the United Kingdom could freely cross the territories of France and Germany again. Most holidaymakers within Europe typically travel across several countries in the EU, and as France and Germany are both popular travel destinations in close geographic proximity, demand shocks likely hit fuel stations in the two countries in a similar way. In addition, we directly account for demand-related shocks by including regional information on the daily mobility to work and to retail and recreational places as control variables into our empirical specification.

Transitory supply shocks should affect French and German fuel stations in a similar way. Due to their geographic proximity, the fuel stations in France and Germany procure most of their crude oil from similar sources. The two countries are also members of the European Single Market, which implies harmonized border checks, common customs policy, and identical regulatory procedures on the movement of goods within the EU.

Finally, no major reforms were implemented in France during our analysis period. In general, there are no fuel price-setting regulations in Germany and France, and both countries have mandatory disclosure of fuel prices, which reaffirms our choice of France as a suitable control group.

\section{Results}\label{sec: results}

This section presents the effects of the temporary VAT reduction on fuel prices and retail margins. We estimate these effects for the three main fuel types: \textit{E5}, \textit{E10}, and diesel.

\subsection{Effect of the VAT Rate Reduction on Prices}

\begin{comment}
Table \ref{tab2} shows the average treatment effects of the consumption tax change on fuel prices, estimated following a regression model presented in Equation 1. Columns (1)-(3) correspond to the tax incidence on \textit{E5}, \textit{E10}, and diesel prices without mobility controls. Columns (4)-(6) show the effects on prices when mobility controls are included. As expected, higher mobility for retail, recreational, and work purposes positively correlates with fuel prices. 
The results in columns (1)-(3) show that the reduction in VAT led to a decline in prices of all fuel products, which is statistically significant at 1 percent level. Including mobility controls does not significantly change the estimates. Among the three fuel product types, diesel experiences the largest decline of 2.08\% in prices in the post-VAT change period. Prices of \textit{E10} gasoline decline by 1.53\%, whereas \textit{E5} gasoline prices drop by 1.01\%.
\end{comment}

Table \ref{tab2} shows the results of estimating the regression model presented in Equation 1 using the logarithm of price as an outcome variable. The coefficients in Columns (1) to (3) correspond to the effect of the temporary VAT rate reduction on \textit{E5}, \textit{E10} and diesel prices without mobility controls. Columns (4) to (6) show the effects on prices when we control for mobility.

The results in Columns (1) to (3) show that the reduction in the VAT led to a decline in prices of all fuel products, which is statistically significant at the 1 percent level and economically significant. The average price for \textit{E5} decreases by $1.05$ percent after the VAT reduction, whilst average prices for \textit{E10} and diesel decrease by $1.55$ and $2.12$ percent, respectively. Including mobility controls does not significantly change the estimates. Reassuringly, however, mobility for retail, recreational, and work purposes positively correlates with fuel prices. When controlling for regional differences in mobility, average prices for \textit{E5}, \textit{E10} and diesel are estimated to decline by $1.01$, $1.53$ and $2.08$ percent, respectively.

Next, we estimate the pass-through rates of the VAT change. Under full pass-through, we expect prices for each fuel product to decrease by about $2.52$ percent.\footnote{With a decrease in the VAT rate from $19$ percent before the reform to $16$ percent after the reform, this is $\frac{1.16-1.19}{1.19}*100 \approx -2.52\%$.} An estimated decline of $2.08$ percent in diesel prices is therefore relatively close to full pass-through. Around $83$ percent of the VAT reduction is passed on to consumers who refuel with diesel. For \textit{E10}, the pass-through rate is $61$ percent. Finally, we estimate that $40$ percent of the VAT decline is passed on to consumers of \textit{E5}. For all fuel products, pass-through of the VAT reduction is fast and relatively high, but incomplete.

\begin{figure}{}
\caption{Effect of the temporary VAT reduction on average prices, E5 \label{fig_e5p}}{}
\label{fig: main}
\begin{minipage}{\textwidth}
\begin{threeparttable}
\begin{center}
\includegraphics[width=1\textwidth]{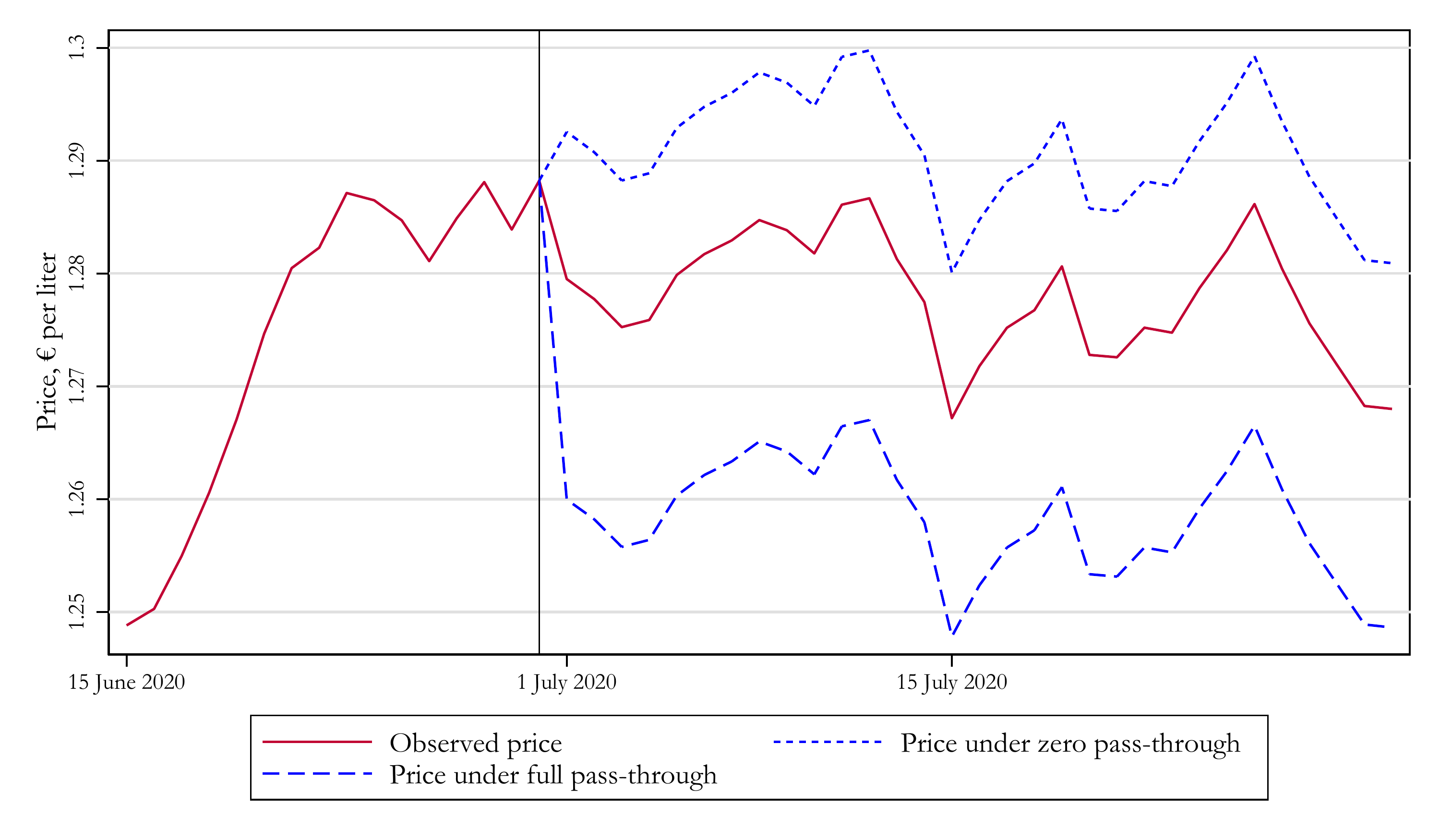}
\end{center}
\begin{footnotesize}
\begin{tablenotes}
\footnotesize Notes: The solid line shows the evolution of the daily weighted average price of \textit{E5} in Germany between 15 June and 31 July 2020. The dashed lines show the counterfactual evolution of the daily weighted average gasoline price in Germany under zero and full pass-through. The solid vertical line shows the beginning of the VAT reduction.
\end{tablenotes}
\end{footnotesize}
\end{threeparttable}
\end{minipage}
\end{figure}

To show the economic importance of the pass-through rate, we illustrate the actual price development, as well as estimated counterfactual prices under full and zero pass-through until 31 July 2020. Figure \ref{fig_e5p} shows these different scenarios for \textit{E5}. The solid red line corresponds to the observed evolution of \textit{E5} prices at German stations from 15 June until 31 July 2020. The vertical line marks the beginning of the post-VAT change period. The long-dashed line corresponds to prices predicted under full pass-through, and the short-dashed line to those under a zero pass-through case.

After the VAT reduction, the observed price of \textit{E5} is at around 1.27 to 1.28 Euro per liter, which is closer to the price that we predict under zero pass-through than to that under full pass-through. The observed \textit{E5} price is on average 2 Eurocent higher than the counterfactual were the retailers to fully pass on the reduction in the VAT rate to consumers.

\begin{figure}{}
\caption{Effect of the temporary VAT reduction on average prices, E10 \label{fig_e10p}}{}
\label{fig: main}
\begin{minipage}{\textwidth}
\begin{threeparttable}
\begin{center}
\includegraphics[width=1\textwidth]{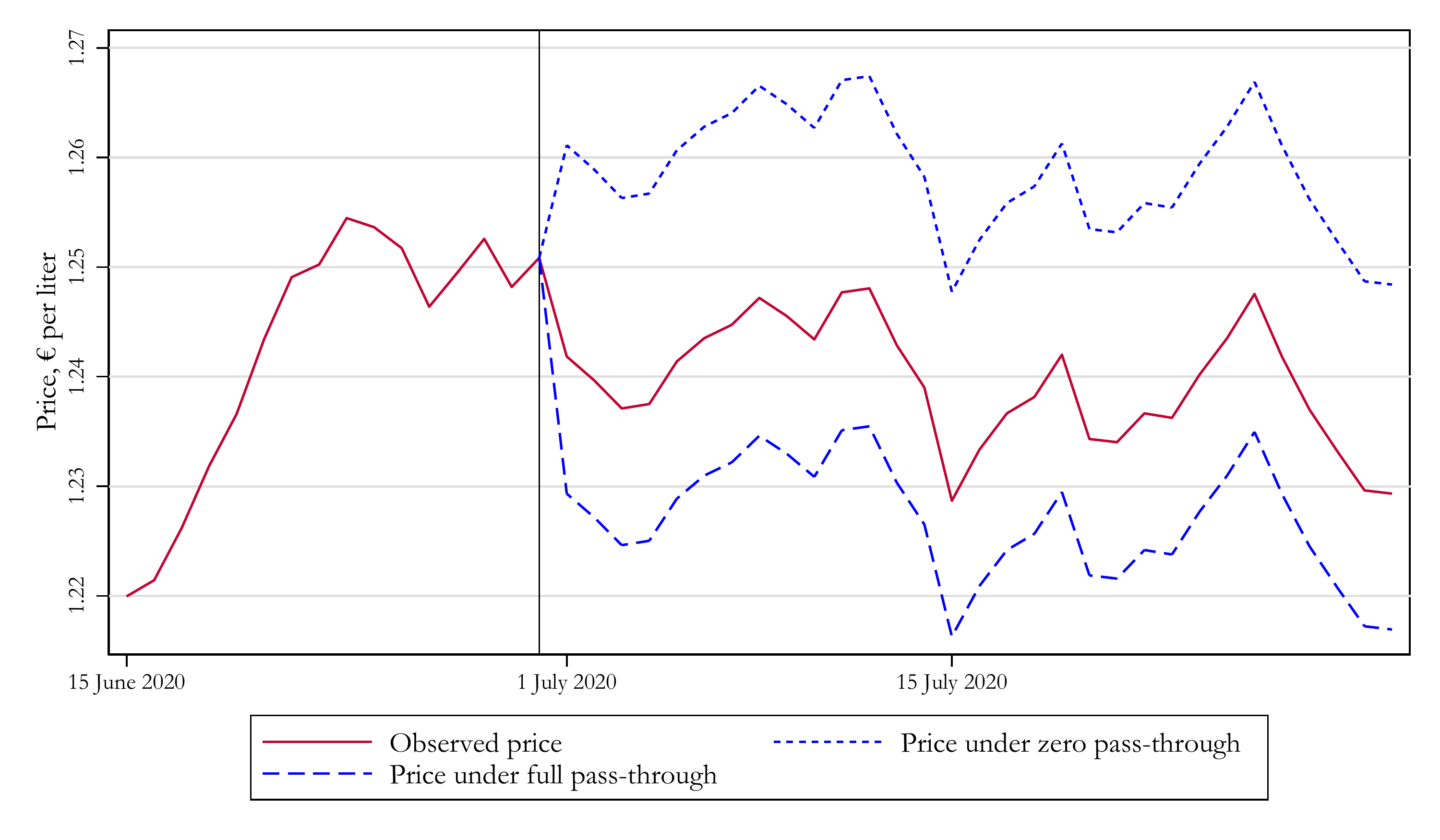}
\end{center}
\begin{footnotesize}
\begin{tablenotes}
\footnotesize Notes: The solid line shows the evolution of the daily weighted average price of \textit{E10} in Germany between 15 June and 31 July 2020. The dashed lines show the counterfactual evolution of the daily weighted average gasoline price in Germany under zero and full pass-through. The solid vertical line shows the beginning of the VAT reduction.
\end{tablenotes}
\end{footnotesize}
\end{threeparttable}
\end{minipage}
\end{figure}

Figure \ref{fig_e10p} shows the VAT incidence for \textit{E10}. The interpretation of the lines is equivalent to Figure \ref{fig_e5p}.

In the post-VAT reduction period the observed \textit{E10} price declines to around 1.23 to 1.25 Eurocent per liter. In the full pass-through case, we predict \textit{E10} prices to be on average 1 Eurocent lower, which is a smaller difference compared to \textit{E5}.

\begin{figure}{}
\caption{Effect of the temporary VAT reduction on average prices, diesel \label{fig_dieselp}}{}
\label{fig: main}
\begin{minipage}{\textwidth}
\begin{threeparttable}
\begin{center}
\includegraphics[width=1\textwidth]{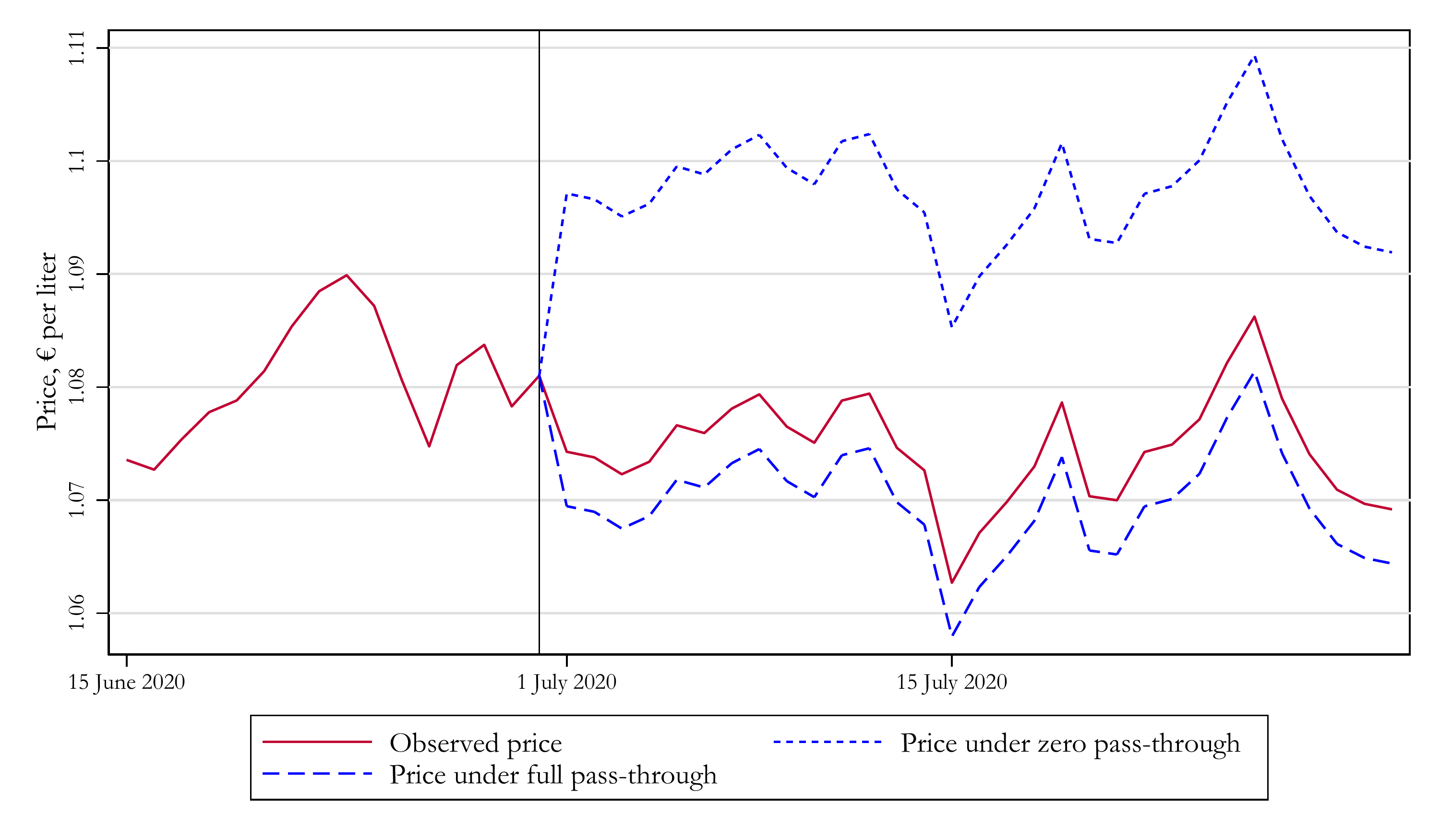}
\end{center}
\begin{footnotesize}
\begin{tablenotes}
\footnotesize Notes: The solid line shows the evolution of the daily weighted average price of diesel in Germany between 15 June and 31 July 2020. The dashed lines show the counterfactual evolution of the daily weighted average diesel price in Germany under zero and full pass-through. The solid vertical line shows the beginning of the VAT reduction.
\end{tablenotes}
\end{footnotesize}
\end{threeparttable}
\end{minipage}
\end{figure}

Figure \ref{fig_dieselp} presents an analogous graph of value-added tax incidence for diesel prices. Among the three fuel products, consumers who fuel with diesel experience the highest pass-through. The observed diesel price in Germany, captured by the solid red line, is relatively close to the price that we predict under the full pass-through scenario. On average, consumers pocket about 2.3 Eurocent per liter of diesel from the VAT reduction.

As we would expect, the ranking of pass-through rates corresponds to the ranking of customer groups with respect to their likelihood of shopping for lower prices. As described in Section \ref{sec: industry}, consumers buying \textit{E5} are those least likely to search for lower prices. Most of these consumers can already save in a similar order of magnitude by switching to \textit{E10}, but choose not to do so. The incentive for fuel stations to pass on the VAT rate reduction to these consumers is therefore also the lowest. In contrast, diesel customers are on average much more likely to be frequent drivers and are therefore much more likely to search for lower prices. Fuel stations therefore experience more competitive pressure for diesel customers and are hence more inclined to pass on the VAT rate reduction to these customers.

\begin{comment}
That a pass-through rate is higher for \textit{E10} than \textit{E5} gasoline is not surprising. Despite a consistently higher average price of \textit{E5} gasoline in Germany, the vast majority of consumers who drive gasoline automobiles still prefer \textit{E5} to \textit{E10}.\footnote{According to the Federal Association of the German Bioethanol Industry, \textit{E5} captured 81.7\% of the gasoline market in 2019.} At the same time, \textit{E5} and \textit{E10} are relatively homogenous products. Only certain older automobiles are not recommended to use \textit{E10} for fuel.\footnote{2016 report by German Automobile Association published that '\textit{E10} was introduced in 2011, to date not a single case of a car using it is known to have been damaged by the fuel.'} Given that around 90\% of gasoline cars can be freely fueled with either \textit{E5} or \textit{E10} gasoline, consumers in Germany appear to display a preference towards \textit{E5}. In turn, a lower price elasticity of demand for \textit{E5} than \textit{E10} gasoline would explain why the retailers choose to pass-through a lower share of the VAT reduction on \textit{E5} price than that of \textit{E10}.

\textit{The pass-through rate is highest for Diesel due to...}

\end{comment}

%\FloatBarrier

\begin{table}\centering
\def\sym#1{\ifmmode^{#1}\else\(^{#1}\)\fi}
\begin{threeparttable}
\caption{Effect of the VAT rate reduction on log prices (percent) \label{tab2}}
\begin{tabular}{l*{6}{c}}
\toprule
                    &\multicolumn{1}{c}{E5}&\multicolumn{1}{c}{E10}&\multicolumn{1}{c}{Diesel}&\multicolumn{1}{c}{E5}&\multicolumn{1}{c}{E10}&\multicolumn{1}{c}{Diesel}\\
\midrule
                    &\multicolumn{1}{c}{(1)}&\multicolumn{1}{c}{(2)}&\multicolumn{1}{c}{(3)}&\multicolumn{1}{c}{(4)}&\multicolumn{1}{c}{(5)}&\multicolumn{1}{c}{(6)}\\
\midrule
VAT reduction       &     -.0105\sym{***}&     -.0155\sym{***}&     -.0212\sym{***}&     -.0101\sym{***}&     -.0153\sym{***}&     -.0208\sym{***}\\
                    &    (.0003)         &    (.0002)         &    (.0002)         &    (.0003)         &    (.0002)         &    (.0002)         \\
\midrule
\addlinespace
Retail \& recreation&                     &                     &                     &      .0001\sym{***}&      .0063\sym{***}&      .0085\sym{***}\\
                    &                     &                     &                     &    (.0000)         &    (.0005)         &    (.0004)         \\
\addlinespace
Workplaces          &                     &                     &                     &      .0001\sym{***}&      .0026\sym{***}&     -.0006         \\
                    &                     &                     &                     &    (.0000)         &    (.0006)         &    (.0005)         \\
\midrule
\addlinespace
Pass-through rate   &          42\%       &          61\%       &          84\%       &      40\%           &       61\%          &       83\%       \\
\midrule
\multicolumn{1}{l}{Date fixed effects}  &\multicolumn{1}{c}{Yes}  &\multicolumn{1}{c}{Yes} &\multicolumn{1}{c}{Yes} &\multicolumn{1}{c}{Yes} &\multicolumn{1}{c}{Yes}    &\multicolumn{1}{c}{Yes}         \\

\multicolumn{1}{l}{Station fixed effects}  &\multicolumn{1}{c}{Yes}  &\multicolumn{1}{c}{Yes} &\multicolumn{1}{c}{Yes} &\multicolumn{1}{c}{Yes} &\multicolumn{1}{c}{Yes}  &\multicolumn{1}{c}{Yes}         \\
\midrule
Observations        &      800,894        &      873,883        &     1,005,583       &      800,382        &      872,763        &     1,004,212         \\
Adjusted \(R^{2}\)  &       0.847         &       0.849         &       0.942         &       0.847         &       0.849         &       0.942         \\
Mean price          &      1.27           &      1.23           &      1.07           &      1.27           &      1.23           &      1.07         \\
\bottomrule
\end{tabular}
\begin{tablenotes}
\footnotesize Notes: Columns (1)-(3) present estimates without mobility control variables on E5, E10, and diesel log prices, respectively. Columns (4)-(6) present estimates on E5, E10, and diesel log prices from estimation with mobility controls. All columns use data from 15 June to 31 July 2020.\\ 
Standard errors clustered at the fuel station level in parentheses.\\
\sym{*} \(p<0.10\), \sym{**} \(p<0.05\), \sym{***} \(p<0.01\)\\
\end{tablenotes}
\end{threeparttable}
\end{table}

%\FloatBarrier

\FloatBarrier

\subsection{Effect of the VAT Rate Reduction on Retail Margins}

\begin{comment}
Table \ref{tab3} presents the effects of the VAT reduction on retail margins, estimated following a regression model in Equation 1. Results in Columns (1)-(3) correspond to the estimation without mobility variables, whereas those in Columns (4)-(6) include controls on mobility. The results in Columns (1)-(3) show that the reduction in VAT rate led to an increase in retail margins at German filling stations that is statistically significant at 1 percent level. Including control variables on mobility for retail, recreational, and work purposes does not significantly alter the estimates. On average, the retail margins increase by .26 Eurocent per liter on diesel, 1.02 Eurocent per liter on \textit{E10}, and by 1.61 Eurocent per liter on \textit{E5} gasoline with the VAT change.
\end{comment}

Table \ref{tab3} presents the results of estimating the regression model presented in Equation 1 using the logarithm of retail margins as an outcome variable. The coefficients in Columns (1) to (3) correspond to the effect of the VAT rate reduction on \textit{E5}, \textit{E10} and diesel margins without controlling for mobility. Columns (4) to (6) show the effects on margins when mobility controls are included.

The results in Columns (1) to (3) show that the reduction in the VAT led to an increase in retail margins for all fuel products, which is statistically significant at the 1 percent level and economically significant. The average retail margin for \textit{E5} increases by $11.34$ percent after the VAT reduction, whilst the average retail margin for \textit{E10} and diesel increases by $10.59$ and $0.46$ percent, respectively. Including mobility controls does not significantly change the estimates. After controlling for regional differences in mobility, average retail margins for \textit{E5}, \textit{E10} and diesel increase by $11.68$, $10.83$ and $0.70$ percent, respectively.

These findings are in line with the estimated pass-through rates. For diesel, where most of the VAT reduction is passed through to consumers, there is only a very modest increase in retail margins. Instead for gasoline, particularly for \textit{E5}, less than half of the VAT rate reduction is passed on to consumers and thus the temporary decrease in the VAT rate led to an increase in retail margins by $10$ to $12$ percent. Note, that for our measure of retail margins we simply subtract taxes and duties, as well as the share of the crude oil price attributable to the production of diesel and gasoline, from the gross price. The retail margins therefore include the refinery margin, as well as the station margin and different cost types, such as the cost of refining or the cost of transportation. Since this means that our estimated retail margin is an overestimate of the actual retail margin, the estimated percentage increase in margins due to the VAT rate reduction can be taken as a lower bound. Thus, the VAT rate reduction likely led to an increase in retail margins of much more than 10 percent.

\begin{table}\centering
\def\sym#1{\ifmmode^{#1}\else\(^{#1}\)\fi}
\begin{threeparttable}
\caption{Tax incidence on log retail margins \label{tab3}}
\begin{tabular}{l*{6}{c}}
\toprule
                    &\multicolumn{1}{c}{E5}&\multicolumn{1}{c}{E10}&\multicolumn{1}{c}{Diesel}&\multicolumn{1}{c}{E5}&\multicolumn{1}{c}{E10}&\multicolumn{1}{c}{Diesel}\\
\midrule
                    &\multicolumn{1}{c}{(1)}&\multicolumn{1}{c}{(2)}&\multicolumn{1}{c}{(3)}&\multicolumn{1}{c}{(4)}&\multicolumn{1}{c}{(5)}&\multicolumn{1}{c}{(6)}\\
\midrule
VAT reduction       &      .1134\sym{***}&      .1059\sym{***}&      .0046\sym{***}&      .1168\sym{***}&      .1083\sym{***}&      .0070\sym{***}\\
                    &    (.0016)         &    (.0016)         &    (.0010)         &    (.0017)         &    (.0017)         &    (.0010)         \\
\addlinespace
Retail \& recreation&                     &                     &                     &      .0779\sym{***}&      .0807\sym{***}&      0.0494\sym{***}\\
                    &                     &                     &                     &    (.0040)         &    (.0039)         &    (.0025)         \\
\addlinespace
Workplaces&                     &                     &                     &      .0458\sym{***}&      .0393\sym{***}&     -.0116\sym{***}\\
                    &                     &                     &                     &    (.0047)         &    (.0045)         &    (.0026)         \\
\midrule
\multicolumn{1}{l}{Date fixed effects}  &\multicolumn{1}{c}{Yes}  &\multicolumn{1}{c}{Yes} &\multicolumn{1}{c}{Yes} &\multicolumn{1}{c}{Yes} &\multicolumn{1}{c}{Yes}    &\multicolumn{1}{c}{Yes}         \\

\multicolumn{1}{l}{Station fixed effects}  &\multicolumn{1}{c}{Yes}  &\multicolumn{1}{c}{Yes} &\multicolumn{1}{c}{Yes} &\multicolumn{1}{c}{Yes} &\multicolumn{1}{c}{Yes}  &\multicolumn{1}{c}{Yes}         \\
\midrule
Observations        &      800,841        &      873,721        &     1,005,559         &      800,329         &      872,601         &     1,004,188         \\
Adjusted \(R^{2}\)  &       0.778         &       0.765         &       0.783         &       0.779         &       0.765         &       0.783         \\
Mean retail margin     &     0.15         &     0.12         &     0.17         &     0.15         &     0.12         &     0.17         \\
\bottomrule
\end{tabular}
\begin{tablenotes}
\footnotesize Notes: Columns (1)-(3) present estimates without mobility control variables on E5, E10, and diesel log retail margins, respectively. Columns (4)-(6) present estimates on E5, E10, and diesel log retail margins from estimation with mobility controls. All columns use data from 15 June to 31 July 2020.\\ 
Standard errors clustered at the fuel station level in parentheses.\\
\sym{*} \(p<0.10\), \sym{**} \(p<0.05\), \sym{***} \(p<0.01\)\\
\end{tablenotes}
\end{threeparttable}
\end{table}

\begin{figure}{}
\caption{Effect of the temporary VAT reduction on average retail margins, E5 \label{fig_e5m}}{}
\label{fig: main}
\begin{minipage}{\textwidth}
\begin{threeparttable}
\begin{center}
\includegraphics[width=1\textwidth]{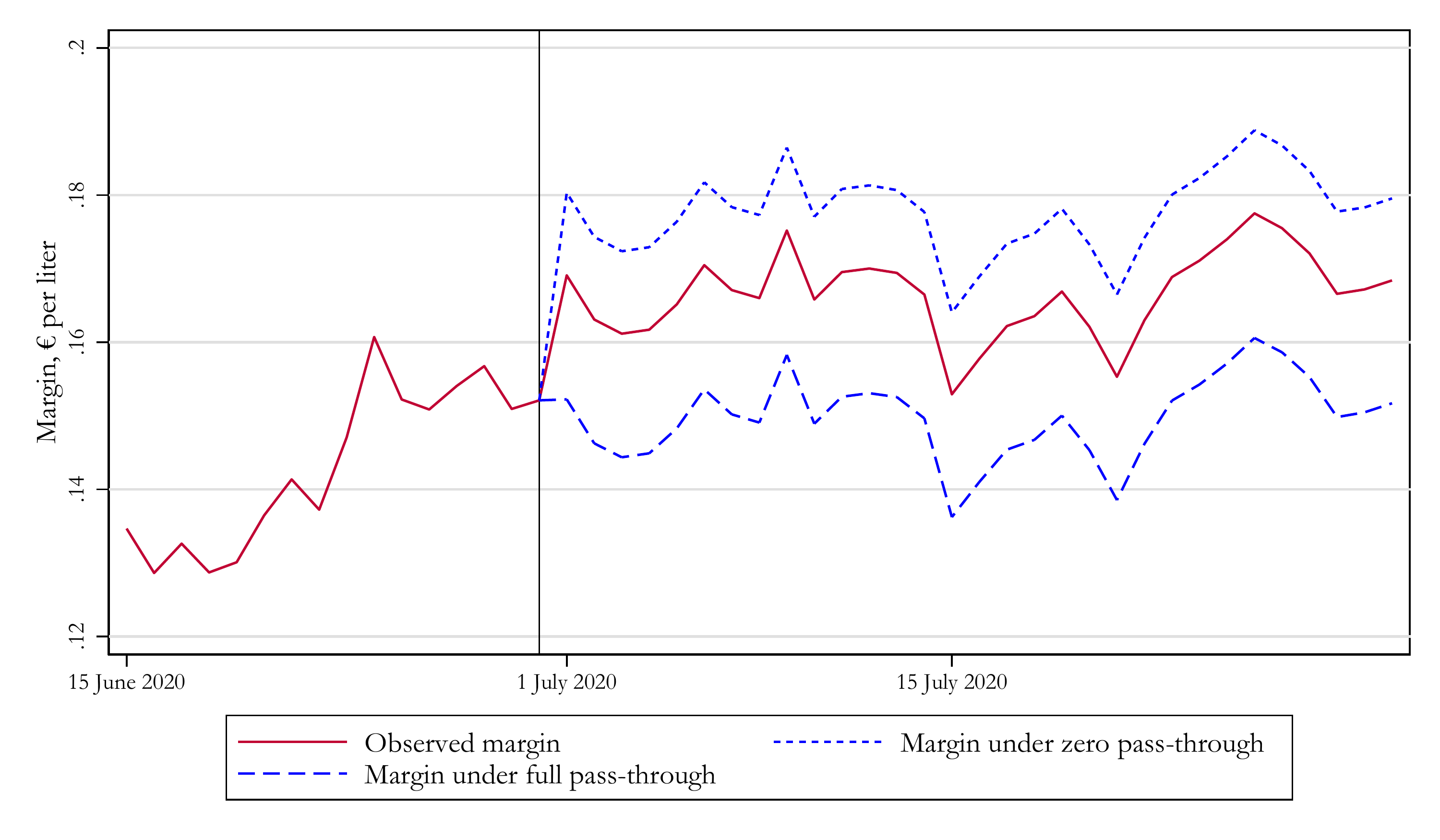}
\end{center}
\begin{footnotesize}
\begin{tablenotes}
\footnotesize Notes: The solid line shows the evolution of daily weighted average retail margins of fuel stations in Germany between June and July 2020. The dashed lines show the counterfactual evolution of the daily weighted average retail margins of fuel stations in Germany under zero and full pass-through. The solid vertical line shows the beginning of the VAT reduction.
\end{tablenotes}
\end{footnotesize}
\end{threeparttable}
\end{minipage}
\end{figure}

Figure \ref{fig_e5m} illustrates the evolution of retail margins for \textit{E5}, as well as the counterfactual under full and zero pass-through. The solid red line corresponds to the observed evolution of retail margins at German fuel stations between 15 June and 31 July 2020. In addition, we plot margins predicted under zero and full pass-through scenarios with short- and long-dashed lines, respectively. For \textit{E5}, the observed evolution of retail margins is closer to the zero pass-through case than to full pass-through.

\begin{figure}{}
\caption{Effect of the temporary VAT reduction on average retail margins, E10 \label{fig_e10m}}{}
\label{fig: main}
\begin{minipage}{\textwidth}
\begin{threeparttable}
\begin{center}
\includegraphics[width=1\textwidth]{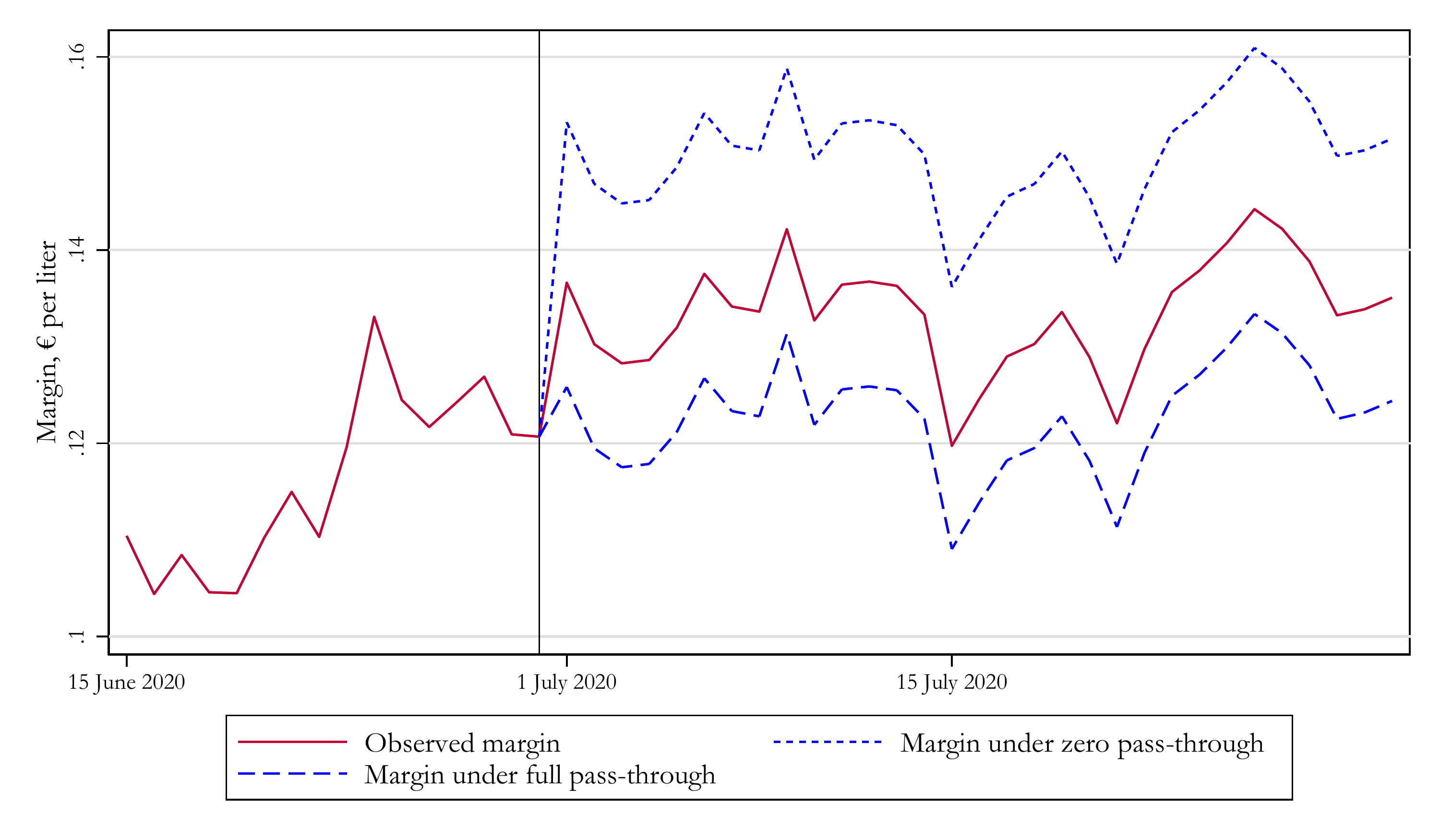}
\end{center}
\begin{footnotesize}
\begin{tablenotes}
\footnotesize Notes: The solid line shows the evolution of daily weighted average retail margins of fuel stations in Germany between June and July 2020. The dashed lines show the counterfactual evolution of the daily weighted average retail margins of fuel stations in Germany under zero and full pass-through. The solid vertical line shows the beginning of the VAT reduction.
\end{tablenotes}
\end{footnotesize}
\end{threeparttable}
\end{minipage}
\end{figure}

\begin{figure}{}
\caption{Effect of the temporary VAT reduction on average retail margins, diesel \label{fig_dieselm}}{}
\label{fig: main}
\begin{minipage}{\textwidth}
\begin{threeparttable}
\begin{center}
\includegraphics[width=1\textwidth]{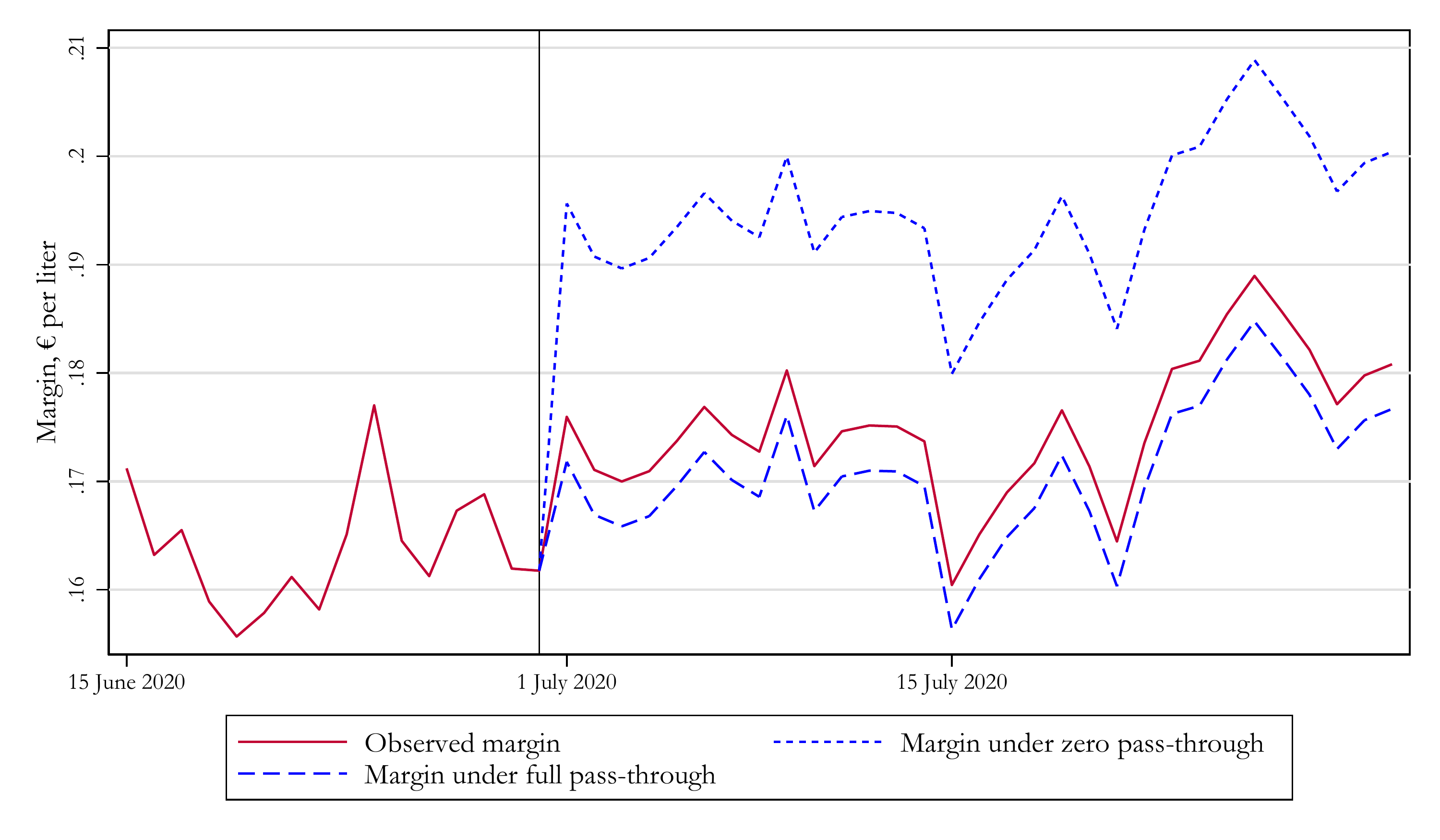}
\end{center}
\begin{footnotesize}
\begin{tablenotes}
\footnotesize Notes: The solid line shows the evolution of daily weighted average retail margins of fuel stations in Germany between June and July 2020. The dashed lines show the counterfactual evolution of the daily weighted average retail margins of fuel stations in Germany under zero and full pass-through. The solid vertical line shows the beginning of the VAT reduction.
\end{tablenotes}
\end{footnotesize}
\end{threeparttable}
\end{minipage}
\end{figure}

Figures \ref{fig_e10m} and \ref{fig_dieselm} present analogous graphs of the evolution of retail margins under full and zero pass-through for \textit{E10} and diesel. In contrast to \textit{E5}, the observed margins of \textit{E10} and diesel are closer to the full pass-through scenario than to zero pass-through.

%\FloatBarrier

\FloatBarrier

%\input{scc_mechanisms}
%this is scc_conclusion

\section{Conclusion}\label{sec: conclusion}

In this paper, we show that, so far, pass-through of the temporary VAT rate reduction, as part of the German fiscal COVID-19 response, is fast and substantial but remains incomplete for all fuel types. Furthermore, we find a high degree of heterogeneity between the pass-through estimates for different fuel types. Since the same stations are selling the different types of fuels and the supply structure is very similar, this leaves differences in competitive pressure from different customer groups as a candidate explanation for the differences in the pass-through rates. In particular, the more likely customers of a particular fuel type are to shop for lower prices the higher is the pass-through rate for this fuel type.

Although fuel markets are not the prime target of unconventional fiscal policy, the mechanisms behind whether and to what extent firms pass on taxes to consumers are the same as for other markets. Studying how fuel stations pass on the temporary VAT reduction is thus a worthwhile exercise and informs us about market-wide pass-through.

A key result is that demand characteristics and competitive pressure play a crucial role in how a temporary tax reduction is passed on to consumers. A key takeaway for policymakers is that by targeting competitive markets with high pass-through rates they can increase the cost-effectiveness of unconventional fiscal policy. In ongoing research we investigate these determinants of pass-through rates in more detail.

%%%%%%%%%%%%%%%%%%%%%%%%%%%%%%%%%%%%%%%%%%%%%%%%%%%%%%%%%%%%%%%%%%%%%%%%%%%%%%%%%%%%%%%%%%%%%%%
% REFERENCES
%%%%%%%%%%%%%%%%%%%%%%%%%%%%%%%%%%%%%%%%%%%%%%%%%%%%%%%%%%%%%%%%%%%%%%%%%%%%%%%%%%%%%%%%%%%%%%%
\newpage

%\bibliographystyle{Bibtex/econometrica}
%\bibliographystyle{/00_Repertoire//ecta}
%\bibliography{/00_Repertoire/bibliography}{}
\printbibliography
%%%%%%%%%%%%%%%%%%%%%%%%%%%%%%%%%%%%%%%%%%%%%%%%%%%%%%%%%%%%%%%%%%%%%%%%%%%%%%%%%%%%%%%%%%%%%%%
% Appendix
%%%%%%%%%%%%%%%%%%%%%%%%%%%%%%%%%%%%%%%%%%%%%%%%%%%%%%%%%%%%%%%%%%%%%%%%%%%%%%%%%%%%%%%%%%%%%%%
\newpage
\begin{appendix}

\noindent
{\LARGE \textbf{Appendix}}
\setcounter{section}{0}%\renewcommand{\thesection}{\arabic{section}} % Roman numerals for the sections
%\renewcommand{\thesubsection}{\arabic{subsection}} % Alphabet for subsections
%\titleformat{\section}{\normalsize\bfseries\filcenter}{\thesection.}{1em}{} % Change section titles
%\titleformat{\subsection}{\normalsize\filcenter}{\thesubsection.}{1em}{} % Change subsection titles
%\setcounter{figure}{0}  \renewcommand{\thefigure}{A.\arabic{figure}}
%\setcounter{table}{0} \renewcommand{\thetable}{A.\arabic{table}}
%this is scc_appendix

\section{Appendix to Section \ref{sec: data}: Data and descriptive evidence}\label{app sec: data}

\begin{comment}
\FloatBarrier
\subsection{Retail Margins and Petrol Station Characteristics in Germany}\label{app sec: descriptives}

\begin{figure}
\caption{Distribution of petrol stations across Germany}
\label{app fig: station distribution}
\begin{minipage}{\textwidth}
\begin{threeparttable}
\begin{center}
\includegraphics[width=0.5\textwidth]{01_figures/stations_map.pdf}
\end{center}
\begin{footnotesize}
\begin{tablenotes}
Note: The Figure shows the geographic distribution of petrol stations in Germany.
\end{tablenotes}
\end{footnotesize}
\end{threeparttable}
\end{minipage}
\end{figure}
\FloatBarrier
\end{comment}

We construct the price panel and compute retail margins at fuel stations in France and Germany as follows. For each fuel station in our data set, we observe a fuel price every time it is changed along with a precise time and date stamp of a change. On average, fuel stations in Germany change fuel prices 15 times a day, whereas there is typically one price change a day at French fuel stations. Based on the distribution of price changes at German fuel stations, we construct hourly fuel prices from 6 am until 10 pm for each day between 15 June - 31 July, 2020. For France, since fuel prices do not change frequently over a day we keep a fuel price at 5 pm for our empirical analysis.

\begin{figure}
\caption{Daily fueling patterns (Germany)}
\label{fig: fueling weights}
\begin{minipage}{\textwidth}
\begin{threeparttable}
\begin{center}
\includegraphics[width=0.9\textwidth]{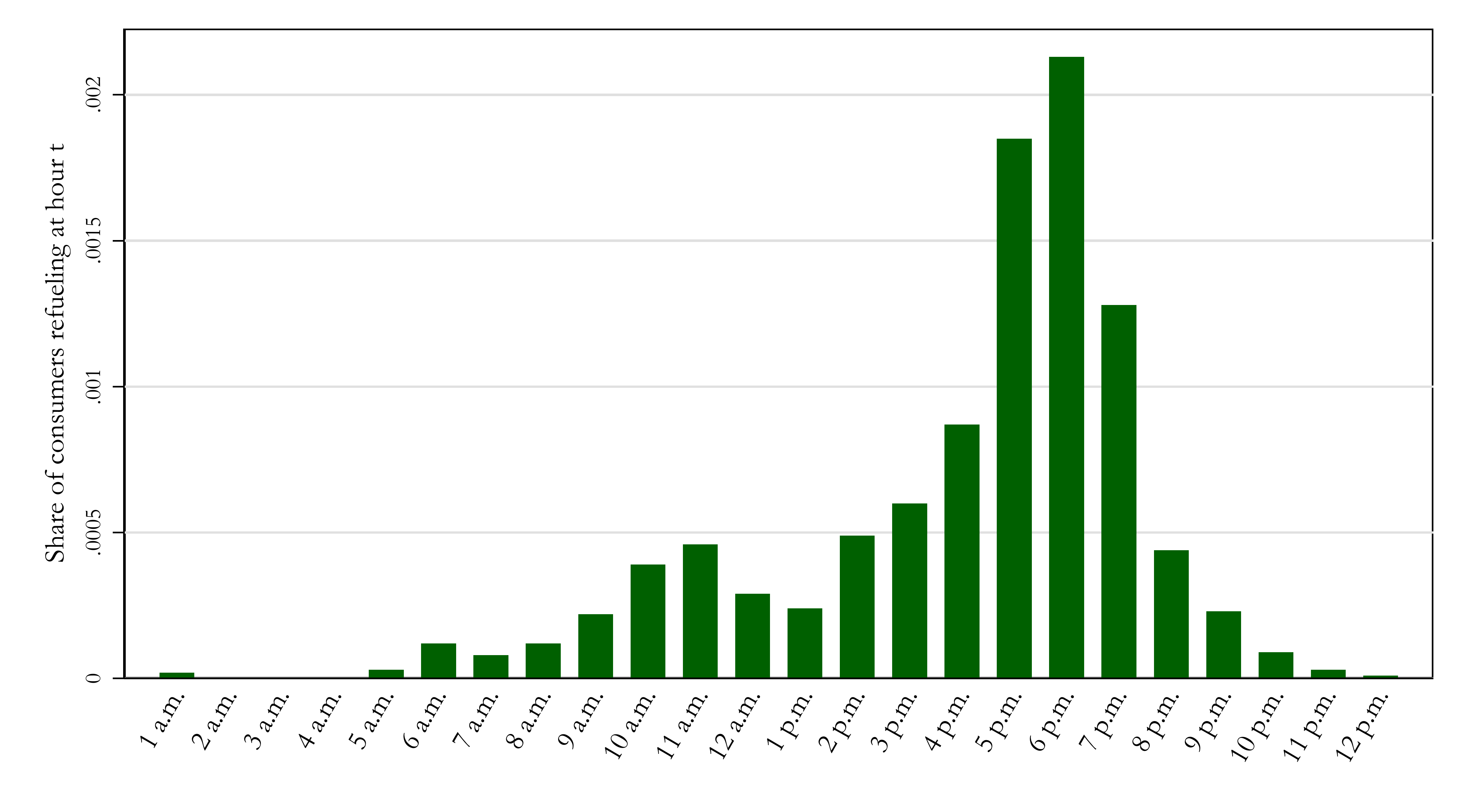}
\end{center}
\begin{footnotesize}
\begin{tablenotes}
Notes: The Figure shows shares of drivers in Germany who fuel at a given hour of a day. Data is based on a representative survey of motorists in Germany, commissioned by the German Federal Ministry of Economic Affairs.
\end{tablenotes}
\end{footnotesize}
\end{threeparttable}
\end{minipage}
\end{figure}

For German fuel stations, we compute daily weighted average price from the hourly distribution of price changes that we observe. To construct the weights, we use the data on hourly fueling patterns reported in a representative survey among drivers by the German Federal Ministry of Economic Affairs. Figure \ref{fig: fueling weights} shows shares of motorists in Germany who fuel at a given time period during a day. We further re-weight the hourly shares to produce weights for the hours between 6 am and 10 pm.

To compute retail margins, we adjust fuel prices to taxes and duties in France and Germany, and subtract a fuel share of the price of crude oil, which is a major input cost. In Germany, taxes and duties consist of the value-added tax, a lump-sum energy tax, and a fee for oil storage. Before the VAT reduction, the value-added tax was at the rate of 19\%, and starting from 1 July 2020 it is temporarily reduced to 16\%. A lump-sum energy tax is at 0.6545 Euro per liter for \textit{E5} and \textit{E10} gasoline, and at 0.4704 Euro per liter for diesel. A fee for oil storage is at 0.27 Euro per liter for \textit{E5} and \textit{E10}, and at 0.30 Euro per liter for diesel.\footnote{See \url{https://www.avd.de/kraftstoff/staatlicher-anteil-an-den-krafstoffkosten/}.}

In France, the value-added tax rate is at 20\%, with the exception of Corsica Island, where it is at 13\%. In addition to the VAT, fuel products in France are subject to a lump-sum tax of 0.60 to 0.70 Euro per liter, depending on a metropolitan region and fuel product type.\footnote{See \url{http://www.financespubliques.fr/glossaire/terme/TICPE/}.} 

We obtain daily data on Brent price of crude oil at the port of Rotterdam from the US Energy Information Administration. A barrel (42 gallons) of crude oil is on average refined into around 19 gallons of gasoline, 12 gallons of diesel, and 13 gallons of other products, such as jet fuel, petroleum coke, and still gas. Among products different from gasoline and diesel, only jet fuel (of which around 4.3 gallons are refined from a barrel of crude oil) yields sizable commercial value. \footnote{See \url{https://www.eia.gov/energyexplained/oil-and-petroleum-products/refining-crude-oil.php}.} 

Assuming that among the other products only jet fuel is of value, we split the price of a barrel into the cost of producing gasoline, diesel, and jet fuel to compute a share of the Brent price that corresponds to a particular fuel product. Around 54\% of the Brent oil price per barrel corresponds to the production of 19 gallons of gasoline, and around 34\% - to the production of 12 gallons of diesel, which we further transform into the input cost per liter of gasoline and diesel. The retail margins of \textit{E5}, \textit{E10}, and diesel are then computed as the observed fuel price adjusted to taxes and duties, minus the share of Brent oil price per liter of a corresponding fuel product.

\FloatBarrier

\end{appendix}
\end{document}